# The dependence of the galaxy mass-metallicity relation on environment and the implied metallicity of the IGM

Ying-jie Peng[1,2], Roberto Maiolino[1,2]

1 Cavendish Laboratory, University of Cambridge, 19 J. J. Thomson Avenue, Cambridge CB3 0HE, UK
2 Kavli Institute for Cosmology, University of Cambridge, Madingley Road, Cambridge CB3 0HA, UK

**ABSTRACT**

We explore the dependence of the galaxy mass-metallicity relation on environment in SDSS, in terms of both over-density and central/satellite dichotomy. We find that at a given stellar mass, there is a strong dependence of metallicity on over-density for star-forming satellites (i.e. all galaxies members of groups/clusters which are not centrals). High metallicity satellites reside, on average, in regions four times denser than the low metallicity ones. Instead, for star-forming centrals no correlation is found. Star-forming satellites at different stellar masses form a tight sequence in the average over-density – metallicity plane, which covers the entire observed range of metallicities and stellar masses. This remarkable result appears to imply that there exists a universal evolutionary path for all star-forming satellites, regardless of their stellar masses. The strong correlation between over-density and metallicity for star-forming satellites indicates that the gas inflow of satellite galaxies is progressively metal-enriched in denser regions. We interpret our results by employing the gas regulator model and find that the metallicity of the enriched inflow of star-forming satellite galaxies, $Z_{0,sat}$, strongly increases with increasing over-density. The derived $Z_{0,sat}$ – overdensity relation is largely independent of stellar mass and can be well described by a simple power law. If the metallicity of the inflow of star-forming satellites can represent the metallicity of the IGM, then the implied metallicity of the IGM rises from $\sim 0.01 Z_\odot$ in the void-like environment to $\sim 0.3 Z_\odot$ in the cluster-like environment, in broad agreement with observations. We show that the observed metallicity difference between star-forming centrals and star-forming satellites becoming smaller towards high stellar masses can be simply explained by the mass-independent enriched inflow, without the need to involve any mass-dependent environmental effect on metallicity. Since satellite galaxies account for at least half of the galaxy population, our findings prompt for a revision of many galaxy evolutionary models, which generally assume pristine gas inflows.

## 1. INTRODUCTION

The stellar mass and environment are two fundamental drivers of the galaxy evolution that have been extensively explored (e.g. Kauffmann et al. 2003 & 2004, Baldry et al. 2006, Peng et al. 2010 (hereafter P10), Peng et al. 2012 (hereafter P12)). In addition to them, metallicity is another crucial diagnostic tool to study the formation and evolution of galaxies. It reflects the metal production and exchange between stars, gas and intergalactic medium (IGM). Therefore, metallicity links together three key processes in galaxy evolution: star formation, gas accretion and galactic outflows (e.g. Davé et al. 2012, Dayal et al. 2012, Lilly et al. 2013). Understanding the interrelationships between stellar mass, environment and metallicity, and their evolution with time is central to understanding the physical processes that control the evolution of the galaxy population.

The mass–metallicity relation for star-forming galaxies has been extensively investigated in the local universe (e.g. Lequeux et al. 1979, Tremonti et al. 2004) and has been extended to higher redshifts (e.g. Savaglio et al. 2005, Erb et al. 2006, Maier et al., 2006, Maiolino et al. 2008, Mannucci et al. 2009, Cresci et al. 2012, Foster et al. 2012, Zahid et al. 2013, Møller et al. 2013). Although the interpretation on the origin of this relation is still not completely clear, it has been well established that metallicity is strongly correlated with galaxy stellar mass. In addition to the mass–metallicity relation, there are many works studying the dependence of this relation on other quantities such as environment (e.g. Shields et al. 1991, Skillman et al. 1996, Mouhcine et al. 2007, Cooper et al. 2008, Ellison et al. 2009, Petropoulou et al. 2011, Pasquali et al. 2012, Hughes et al. 2012), star formation rate (SFR) (e.g. Mannucci et al. 2010, Yates et al. 2012, Pilyugin et al. 2013, Sanchez et al. 2013, Lara-Lopez et al. 2013a) and gas content (e.g. Hughes et al. 2012, Bothwell et al. 2013, Lara-Lopez et al. 2013b).

Regarding the dependence on environment, many studies did not find a significant dependence of the mass–metallicity relation on environment, especially once the degeneracy between mass and environment is properly accounted for. However, most studies do not distinguish between central galaxies (in groups or clusters) and satellites (i.e. all galaxies members of groups/clusters which are not centrals). Because of obvious luminosity and sensitivity biases, the statistics in past studies have been dominated by the properties of the central galaxies, which likely follow evolutionary processes different with respect to the rest of the galaxy population (satellites) in the same over dense region. Pasquali et al. (2012) have differentiated between central and satellite galaxies and have found an interesting, significant offset in terms of metallicity between satellites and centrals at a given stellar mass. However, they do not investigate further this effect as a function of environment over-density. For a given galaxy group, all the group members have the same dark matter halo mass, but they may have very different over-densities that trace primarily the location within the group/halo (see Figure 5 in P12). Therefore, using halo mass as the environment indicator will average over a wide range of over-densities and hence weaken any underlying dependence of the galaxy properties on environment (as traced by the over-density).

In this paper we use the SDSS DR7 (Abazajian et al. 2009) data to further investigate the interrelationships between stellar mass, environment and metallicity. Given the close correlation between mass and metallicity, the primary goal of this paper is therefore to study the dependence of the mass-metallicity relation on environment, in terms of *both* over-density and central/satellite dichotomy. We will show that by disentangling these different properties, tight correlations emerge, which shed light on galaxy chemo-evolutionary scenarios.

Throughout the paper we use a concordance ΛCDM cosmology with $H_0 = 70$ km s$^{-1}$ Mpc$^{-1}$, $\Omega_\Lambda = 0.75$, and $\Omega_m = 0.25$. We use the term "dex" to mean the antilogarithm, i.e., 0.1 dex = $10^{0.1}$ = 1.259.

## 2 SAMPLE AND METHODOLOGY

The sample of galaxies analyzed in this paper is the same SDSS DR7 sample that we constructed and used in P10 and P12. Briefly, it is a magnitude-selected sample of galaxies in the redshift range of $0.02 < z < 0.085$ that have clean photometry and Petrosian SDSS r-band magnitudes in the range of $10.0 < r < 18.0$ after correcting for Galactic extinction. The parent photometric sample contains 1,579,314 objects after removing duplicates, of which 238,474 have reliable spectroscopic redshift measurements. As a consequence of the relatively broad redshift range, the projected physical aperture of the SDSS spectroscopic fiber changes significantly across the sample, however we have verified that the results do not change significantly by selecting narrower redshift ranges. Each galaxy is weighted by 1/TSR× 1/$V_{max}$, where TSR is a spatial target sampling rate, determined using the fraction of objects that have spectra in the parent photometric sample within the minimum SDSS fiber spacing of 55 arcsec of a given object. The $V_{max}$ values are derived from the *k*-correction program v4_1_4 (Blanton & Roweis 2007). The use of $V_{max}$ weighting allows us to include representatives of the galaxy population down to a stellar mass of about $10^9$ M$_\odot$.

The stellar masses are determined from the k-correction code with Bruzual & Charlot (2003) population synthesis models and a Chabrier IMF. When needed, the SFRs of the SDSS star-forming galaxies were derived by Brinchmann et al. (2004, hereafter B04), based on the Hα emission line luminosities, corrected for extinction using the Hα/Hβ ratio, and corrected for fiber aperture effects. The B04 SFR was computed for a Kroupa IMF and so we convert these to a Chabrier IMF, by using log SFR(Chabrier)=log SFR (Kroupa)−0.04. We constrain our analysis to only star-forming galaxies with the classification flag in B04 SFR catalogue (i.e. the "iclass" keyword) set to 1. This also excludes those objects hosting an active galactic nucleus (AGN), for which the derived SFRs and metallicities are probably not reliable.

As in P10 and P12, we characterize the environment of a given galaxy by a dimensionless density contrast over-density δ and by whether the galaxy is a central or a satellite. The group catalog that we use in this paper is the SDSS DR7 group catalog kindly made available by Yang et al. This is the updated version of the Yang et al. SDSS DR4 group catalog, described in Yang et al. (2005, 2007). As in P12, we define central galaxies to be those galaxies that are the most massive and most luminous galaxies within their dark matter halos. Other galaxies lying within the same dark matter halo are defined to be satellites. This operational definition of central eliminates a small fraction (2.1%) of galaxies that would have been classified as centrals using only a mass or luminosity criterion on its own. Including these ambiguous galaxies in the set of centrals produces indistinguishable changes to the results presented in this paper, and is thus not of great importance.

The oxygen gas-phase abundances were measured from the emission line ratios derived by Tremonti et al. (2004, hereafter T04) from SDSS DR7. Following the same SNR thresholds for emission lines as in T04, we restrict our star-forming sample to galaxies that have lines of Hα, Hβ and [N$_{II}$] λ6584 detected at greater than a 5σ level. We have tested other independent measurements of metallicity such as the Mannucci et al. (2010) and Maiolino et al. (2008) metallicity and also find very small changes to the results presented in this paper. The same results are obtained by also using other metallicity calibrations such as the one given in Pettini & Pagel (2004). Obviously, each strong-line metallicity calibration and method gives a different metallicity scale and a different slope of the mass-metallicity relation (Kewley & Ellison 2008), but the trends that are illustrated in the following are preserved regardless of the adopted calibration.

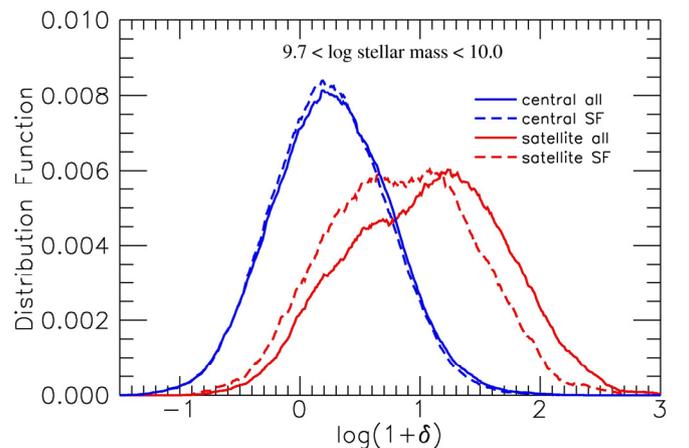

Figure 1. Normalized distribution of centrals and satellites in over-density within a narrow range of stellar mass. Solid lines represent all centrals (in blue) and all satellites (in red). Dashed lines represent star-forming centrals (in blue) and star-forming satellites (in red) with reliable SFRs and metallicities measurements.

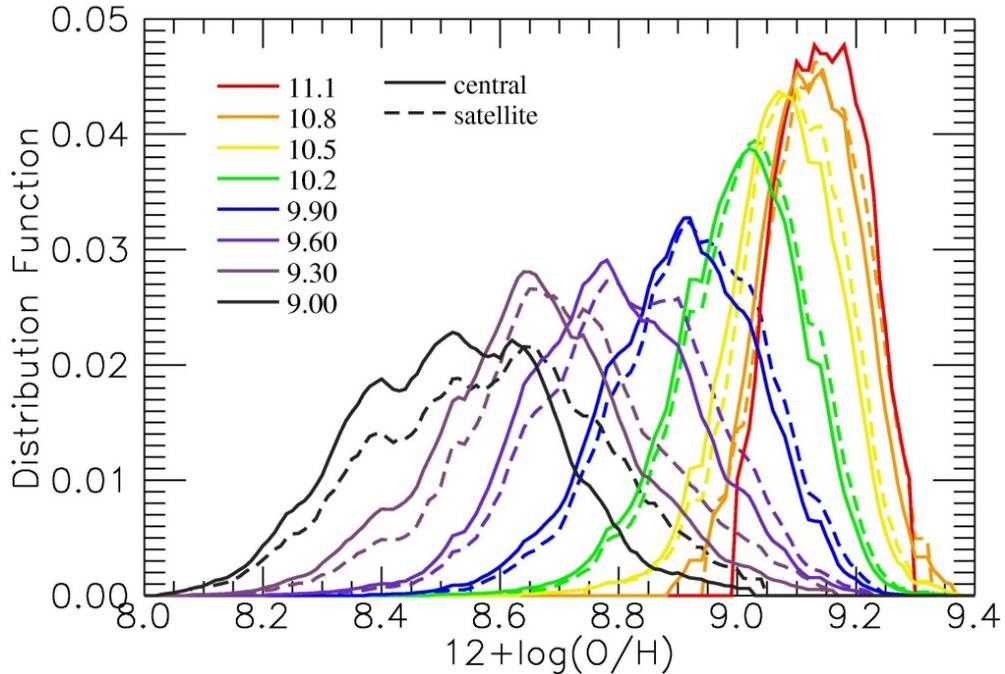

Figure 2. Metallicity distribution for star-forming centrals (solid lines) and star-forming satellites (dashed lines) at different stellar mass bins. At the same stellar mass, the average metallicity of satellites is evidently higher than that of centrals, especially at low stellar masses. The difference of the metallicity distributions for centrals and satellites gradually disappears at high stellar masses.

With all above sample selection criteria, we arrive at a final star-forming galaxy sample including 71,337 objects, of which 15,971 are satellites. All of these objects have reliable over-densities, SFRs, metallicities measurements and unambiguous central/satellite classifications according to our definitions of central and satellite above.

Figure 1 shows the normalized distribution of centrals and satellites in over-density within a (representative) narrow range of stellar mass of $10^{9.7} M_\odot < m_{star} < 10^{10} M_\odot$. By showing a narrow range of stellar mass we eliminate any mass related effect. We have repeated this exercise for other different stellar mass bins and find very similar results, except for the most massive galaxies as a consequence of the fact that the majority of the most massive galaxies are passive central galaxies living in very dense regions. It is interesting to notice in Figure 1 that the normalized distribution of all centrals and that of star-forming centrals are almost exactly the same. While the normalized distribution of star-forming satellites is clearly biased towards lower over-density compared to that of all satellites. This is expected from P12, where we have shown that at a given stellar mass, the red fraction of centrals is largely independent of over-density, but the red fraction of satellites strongly increases with increasing over-density. Therefore, our sample selection effect (to select mainly star-forming galaxies) on centrals should be largely independent of over-density and the selection of star-forming satellites is strongly biased towards lower over-density. However, the distribution of star-forming satellites that satisfy our selection criteria still spans the full range of environments as the full satellite sample probes. Likewise, the fact that normalized distribution of all centrals and that of star-forming centrals are identical can be regarded as the evidence of that the star formation status of centrals is largely independent of environment. In other words, the star-forming centrals with strong emission lines do not have any preference in environment.

## 3 ANALYSIS AND RESULTS

Before discussing the results, we wish to emphasize and clarify that at a given stellar mass, satellites and centrals obviously probe different environments, as shown in Figure 1. More specifically, by definition, a satellite galaxy with a given stellar mass of $M_0$ is hosted in an environment whose central galaxy is more massive than $M_0$; while a central galaxy with the same stellar mass of $M_0$ lives in an environment with satellites less massive than $M_0$. The population of centrals and satellites in the narrow mass range shown in Figure 1 do not belong to the same physical large scale structures. We also make clear that the most massive central galaxies are not included in this study, since these are generally passive ellipticals with no star formation in the local universe, hence are excluded by our requirement of having nebular lines to measure the gas metallicity.

In Figure 2 we show the metallicity distribution for star-forming centrals and star-forming satellites in different stellar mass bins. The dependence of metallicity on stellar mass (mass-metallicity relation) is clearly visible for both star-forming centrals and star-forming satellites, as the metallicity distribution gradually shifts to higher values with increasing stellar masses. At the same stellar mass, the average metallicity of satellites is evidently higher than that of centrals, especially for low stellar mass galaxies. The difference of metallicity distribution for centrals and satellites gradually disappears towards high stellar masses, in agreement with Pasquali et al (2012).

The fact that the dispersion of the metallicity distribution becomes narrower and the peak of the distribution shifts towards the high metallicity with increasing stellar mass suggest that the metallicity is getting progressively saturated with increasing stellar mass at around the yield of ~ 9.1 in units of 12+log(O/H). This saturation effect is consistent with Zahid et al. (2013), who explored the evolution in the shape of the mass-metallicity relation as a function of redshift and found that the mass-metallicity relation flattens at late times. They argue that there exists an empirical upper limit to the metallicity in star-forming galaxies that is independent of redshift. As this saturation effect, which is presumably due to the existence of an upper limit of the metallicity as found by Zahid et al. (2013), is not the key point of the current paper, we will discuss it in our future work.

Given the qualitative similarity of the metallicity distributions for centrals and satellites, we further investigate the role of environment (in terms of over-density) on the mass–metallicity relation. As we have argued in introduction, for the purpose of studying the dependence of galaxy properties on environment, in general over-density is a better indicator of the environment than the dark matter halo mass of the parent group; indeed for a given galaxy group, all the group members have the same halo mass, but they may have very different over-densities that trace primarily the location within the group/halo (see the lower panel of Figure 5 in P12).

For a given set of star-forming galaxies at fixed stellar masses on the overdensity-metallicity plane, there are two complementary ways to study the correlation between over-density and metallicity. The first way is to plot the average metallicity as function of over-density for star-forming centrals and star-forming satellites in different stellar mass bins, as shown in the top panel of Figure 3. These are determined within a moving slide of size 0.3 dex in over-density. We choose narrow stellar mass bin of size 0.2 dex to eliminate the strong dependence of metallicity on stellar mass. The hatched regions around lines show the standard deviation of the sample in that bin. The horizontal dashed lines show the average metallicity of the star-forming centrals at a given stellar mass. For sake of clarity, we show only one stellar mass bin every two, to avoid heavy overlaps between lines from different stellar mass bins.

The second way is to plot the average over-density as a function of metallicity for star-forming centrals and star-forming satellites in different stellar mass bins, as shown in the lower panel of Figure 3. These are determined within a moving slide of size 0.2 dex in metallicity. We use the same narrow stellar mass bin of size 0.2 dex to eliminate the strong dependence of metallicity on stellar mass. The gray hatched regions around lines show the standard deviation of the sample in each bin.

It should be noted that the two panels in Figure 3 are obviously intrinsically related to each other, since both of them represent the same underlying distribution of the star-forming centrals and satellites in the overdensity-metallicity plane. For instance, at a given stellar mass, if the average metallicity of star-forming centrals is independent of over-density, it essentially requires the average over-density of centrals to be independent of metallicity, and vice versa.

Remarkably, both panels of Figure 3 clearly show that, at a given stellar mass, there is a strong correlation between the metallicity and over-density for star-forming satellites. At a given stellar mass, the average metallicity of satellites steadily increases with over-density (top panel) and, vice versa, the average over-density steadily increases with metallicity (lower panel). For the star-forming centrals, interestingly, no significant correlation is found between metallicity and over-density, i.e. all the lines of centrals are essentially flat in both panels. These results clearly demonstrate the important role of environment in shaping the metallicity of the galaxies. At a given stellar mass, star-forming satellites living in over dense regions tend to have higher metallicities while the ones living in under dense regions tend to have lower metallicities. For star-forming centrals, the environment has evidently no effect on their metallicity (once the mass-environment degeneracy is accounted for, i.e. at fixed stellar mass).

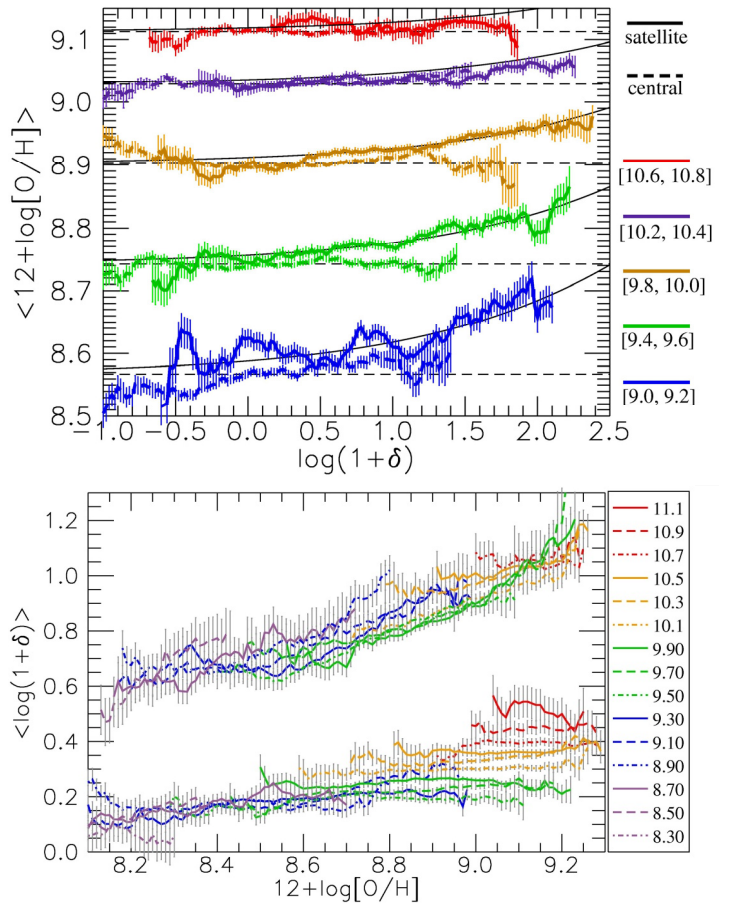

Figure 3. Top panel: The average metallicity as function of over-density for star-forming centrals (dashed lines) and star-forming satellites (solid lines) for different stellar masses. Horizontal black dashed lines show the average metallicity of the star-forming centrals at a given stellar mass. Solid black lines correspond to the simple parameterizations given in the text in Equation (2). For sake of clarity, we show only one stellar mass bin every two, to avoid heavy overlaps between lines from different stellar mass bins. Lower panel: The average over-density as function of metallicity for star-forming centrals (lower set of lines) and star-forming satellites (upper set of lines) for different stellar masses.

Looking more in detail at Figure 3, first we recall that, at a given stellar mass, satellites and centrals probe different environments as shown in Figure 1. Therefore, in the top panel of Figure 3, at a given stellar mass the average metallicity of star-forming centrals probes relatively low density regions while that of satellites probes relatively high density regions. Second, at a given over-density, the difference of the metallicity between the star-forming centrals and star-forming satellites evidently becomes smaller and eventually disappears with increasing stellar mass (top panel of Figure 3). We will show in Section 4.2 that this *mass-dependent* metallicity difference can be simply explained by a *mass-independent* metal-enriched inflow (Equation 2 in the next section), shown by the black solid lines in the top panel of Figure 3, which trace well the observed average metallicity of the star-forming satellites at all stellar masses.

Turning to the lower panel of Figure 3, different lines that represent different stellar mass bins cover their corresponding ranges of the metallicity distribution as shown in Figure 2. For star-forming satellites, all individual line ramps up with increasing metallicity, except for the most massive galaxies at $m_{star} > 10^{11} M_\odot$ for which the metallicity is nearly saturated at around the yield. When we combine all the individual lines together, remarkably, they form a very tight sequence in the average over-density – metallicity plane, spanning the entire observed range of metallicities. In other words, there exists a tight sequence for the star-forming satellites in the average over-density – metallicity plane, *largely independent of the stellar masses*.

When looking at star-forming centrals, first, all individual lines are essentially flat, implying no correlation between environment and metallicity at a given stellar mass. Second, the mass segregation effect for star-forming centrals is clearly visible, i.e. more massive centrals on average live in denser regions.

## 4 DISCUSSION

### 4.1 Dependence of galactic metallicity on environment

First we stress again that the tight sequence of the over-density – metallicity relation for star-forming satellites as shown in the lower panel of Figure 3 is not due to the strong dependence of metallicity on stellar mass. As clearly shown in the same panel, all individual lines (i.e. at each fixed stellar mass) ramp up with increasing metallicity (except for the most massive galaxies at $m_{star} > 10^{11} M_\odot$ due to the metallicity saturation effect). While for the star-forming centrals, despite of the mass segregation effect that shifts the lines upwards vertically with increasing stellar mass, all individual lines are essentially flat which implies there is no correlation between environment and metallicity at a given stellar mass.

To better understand the metallicity dependence on environment at a given stellar mass as shown in Figure 3, we employ the gas regulator model presented in Lilly et al (2013). This simple analytical physical model takes into account the key physical processes of inflow, star formation, outflow and metal production. The formation of stars is instantaneously regulated by the mass of gas reservoir with mass-loss scaling with the SFR. This model has successfully reproduced many key features of the galaxy population, such as the fundamental metallicity relation (FMR) found by Mannucci et al. (2010). Similar models have been built and presented in many works (e.g. Finlator et al. 2008, Recchi et al. 2008, Davé et al. 2012, Dayal et al. 2013). The difference between different models lies in the assumptions made in the construction of the models. For instance, in Davé et al. (2012) the gas mass is assumed to be constant with epoch, as they argue that star-forming galaxies in hydro-dynamic simulations are usually seen to lie near the equilibrium condition; in Dayal et al. (2013) the gas inflow rate is assumed to be proportional to the SFR in order to obtain the simplest analytical solution.

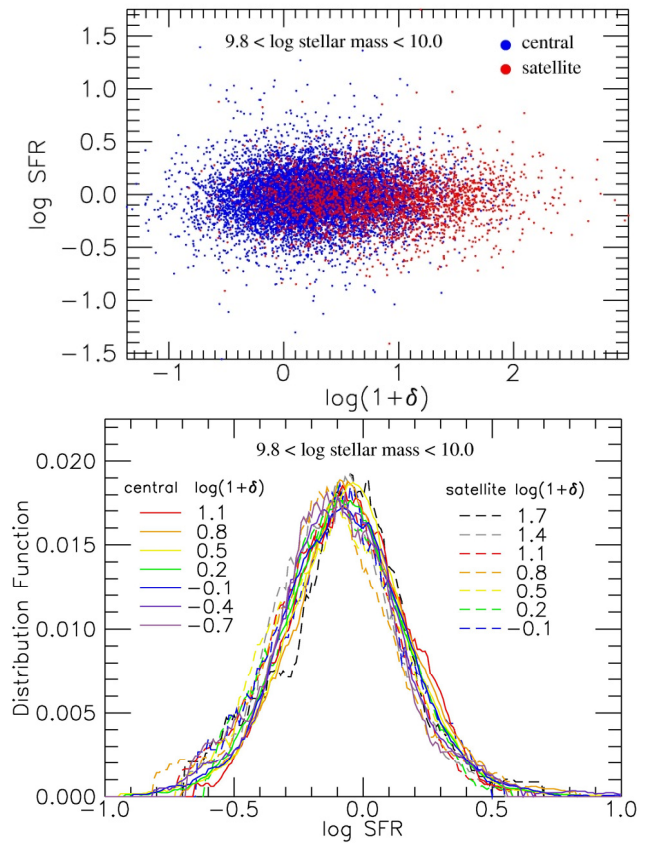

Figure 4. Top panel: SFR and over-density for star-forming centrals and star-forming satellites within a narrow range of stellar mass. By limiting the range of stellar mass we eliminate any mass related effect. Lower panel: the normalized SFR distribution for star-forming centrals and star-forming satellites at different over-density bins within the same limited range of stellar mass. It clearly shows that the normalized SFR distribution function at a given stellar mass is universal and depends neither on over-density nor on the central/satellite dichotomy.

In Lilly et al (2013), the equilibrium metallicity in their gas regulator model is given by

$$Z_{eq}(m_{star}, \rho) = Z_0(m_{star}, \rho) + y(m_{star}, \rho) \frac{SFR(m_{star}, \rho)}{\Phi(m_{star}, \rho)} \quad (1)$$

where $Z_{eq}$ is the equilibrium metallicity, $Z_0$ is the metallicity of the infalling gas, $y$ is the yield, $\Phi$ is the gas inflow rate and $\rho$ indicates the dependence of all quantities on the local over-density of galaxies (environment). The explicit dependence of $Z_{eq}$ on the outflow is hidden in the outflow dependence of SFR/$\Phi$, since in the equilibrium state galaxies reach a dynamical balance between inflow, SFR and outflow. SFR is linked directly to the outflow via the mass-loading factor. Not to lose generality, we first assume that all the parameters in equation (1) may dependent on environment, in terms of over-density and central/satellite dichotomy.

Figure 4 shows the star formation rate in central and satellite galaxies as a function of galaxy over-density, within a limited narrow (representative) range of stellar mass of $10^{9.8} M_\odot < m_{star} < 10^{10} M_\odot$. Figure 4 illustrates that within a limited narrow range of stellar mass, the SFR distribution function of star-forming galaxies does not depend on over-density and central/satellite dichotomy. We have repeated this exercise with other different stellar mass bins and find very similar result. Therefore, at a given stellar mass, the SFR of star-forming galaxies is universal and independent of environment. If we assume that at a given stellar mass, the yield is also independent of environment, then Equation (1) can be simplified to

$$Z_{eq}(m_{star},\rho) = Z_0(m_{star},\rho) + y(m_{star})\frac{SFR(m_{star})}{\Phi(m_{star},\rho)} \quad (2)$$

Equation (2) indicates that at a given stellar mass, the galactic metallicity is proportional to the metallicity of the infalling gas and inversely proportional to the gas inflow rate.

For star-forming satellites, the strong dependence of metallicity on environment at a given stellar mass is then either due to metal-enriched inflow in dense regions (i.e. $Z_0$ is proportional to over-density), or a suppressed gas inflow rate in dense regions (i.e. $\Phi$ is inversely proportional to over-density), or both. The argument on $Z_0$ is expected, as by definition there are more galaxies in dense regions and therefore more metals are produced and expelled into the IGM. This will enhance the $Z_0$ in over dense regions. On the other hand, for a given galaxy group, the over-density is inversely proportional to the distance to the central as shown in Figure 5 of P12. The central by definition is the most massive galaxy in its group and hence injects more metals into the IGM than the satellites surrounding it. This becomes especially true for fossil groups with a large luminosity gap, where the central is much more massive than its satellites. Therefore for a given group, a higher value of over-density of a satellite means it is closer to the central and hence its gas inflow is more likely and heavily to be metal-enriched. The satellites themselves also obviously contribute to the enrichment of the IGM through their star formation-driven outflows, and this process is also proportional to the number density of satellite galaxies (i.e. the overdensity).

As for the argument on $\Phi$, a suppressed gas inflow rate in dense regions is indeed able to produce the dependence of the metallicity on environment for star-forming satellites via Equations (2). However, at a given stellar mass, if the star-formation efficiency is independent of environment, a reduced inflow rate will tend to deplete the gas reservoir and thus to suppress the SFR, which would be in contrast to the universal SFR distribution as shown in Figure 4.

Turning to star-forming centrals, there is no correlation found between environment and metallicity at a given stellar mass, which is clearly different from the behavior of the star-forming satellites. Similar to the $Z_0$ argument as discussed above, this result can be explained by the simple fact that the central galaxy is the most massive one in its group by definition. As a consequence, for a given group, the central galaxy has the highest metallicity and also produces more metals than the satellites surrounding it. This becomes especially true for fossil groups with a large luminosity gap. Therefore it is expected that the metallicity enhancement of the satellites that surround the central, due to the metals produced and expelled by the central into the IGM, is much stronger compared to the metallicity enhancement of the central, due to the metals produced and expelled by its satellites. On the other hand, for a given group, in order to enhance the metallicity of the central, it needs relatively higher metal-enriched inflow than the satellites, as the central galaxy has the highest metallicity in the group. This mass-dependent metallicity enhancement due to the metal-enriched inflow will be discussed more in detail and more quantitatively in next section.

In addition to the mass-dependent metallicity enhancement effect discussed above, the fact that the metallicity of the central does not depend on environment could also be a consequence of them lying at the bottom of the halos' potential well and fed directly by the cold streams that transport material to the center of the halo, according to the scenario presented by Dekel et al. (2009). Therefore, in this scenario, the infalling gas towards the centrals may likely be pristine and may have not been metal enriched.

Our results are qualitatively in good agreement with the predictions from cosmological hydrodynamic simulations of Davé et al (2011). They found in their simulations that galaxies in denser regions and satellites have higher metallicities, since these systems are obtaining more enriched inflow. Oppenheimer et al (2012) show in their simulations that the metals in the IGM are progressively more abundant in more over-dense regions.

Finally, we emphasize that the tight over-density – metallicity relation for star-forming satellites, covering the entire observed metallicity range, is *independent of stellar mass*. This surprising result suggests the existence of a universal evolutionary path for all star-forming satellites. A given star-forming satellite galaxy will increase its stellar mass via star formation, migrate to higher over-density via structure growth in the universe and also accordingly increase its metallicity in dense regions. If individual star-forming satellite galaxies with different stellar masses follow different over-density increase history and/or different metallicity enrichment history, their evolutionary trajectories on the average over-density – metallicity plane should be highly mass-dependent. Therefore, the fact that we find a tight sequence that is largely independent of stellar mass suggests that the majority of the star-forming satellites may follow very similar density increase history and connected metallicity enrichment history. These three processes, i.e. star formation, structure growth and metallicity enhancement, are intrinsic processes of the star-forming satellites and they are working together to produce this tight sequence for the star-forming satellites. However, we cannot exclude that some other physical mechanism or several processes may conspire together to produce such a tight sequence.

### 4.2 Metallicity of the inflow and IGM

In the previous section we have qualitatively shown that the observed dependence of the metallicity on environment for star-forming satellites can be explained by metal-enriched inflows in dense regions. In this section we quantitatively calculate what the metallicity of the infalling gas needs to be in order to reproduce the observations, by employing the gas regulator model and Equation (2).

Since the metallicity of the IGM in the void-like environment is observed to be below $0.02 Z_\odot$ (Stocke et al. 2007) and the metallicity of star-forming centrals is independent of environment (Figure 3), we can assume that the inflow of star-forming centrals is about pristine in all environments (if this happens through direct cold flows from the outer halo), i.e. assume the metallicity of the inflow of star-forming centrals $Z_{0,cen} \sim 0$ (however, as it will be discussed later on, this assumption is not crucial to explain the metallicity behavior of centrals). In Figure 5, the observed average metallicity of star-forming centrals is plotted as the black line. We adopt the solar oxygen abundance of 8.69 (Asplund et al. 2009), in units of 12+log(O/H). The metallicity enhancement due to the enriched inflow $Z_0$ can be easily derived via Equation (2). As we have discussed before, at a given stellar mass the SFR distribution function is universal and depends neither on over-density nor on central/satellite dichotomy. As discussed in the previous section, as a consequence of the universal SFR distribution, the inflow rate $\Phi$ is also unlikely to depend on over-density or on central/satellite dichotomy. Therefore, $Z_0$ is likely to be the most environment-dependent quantity in Equation (2).

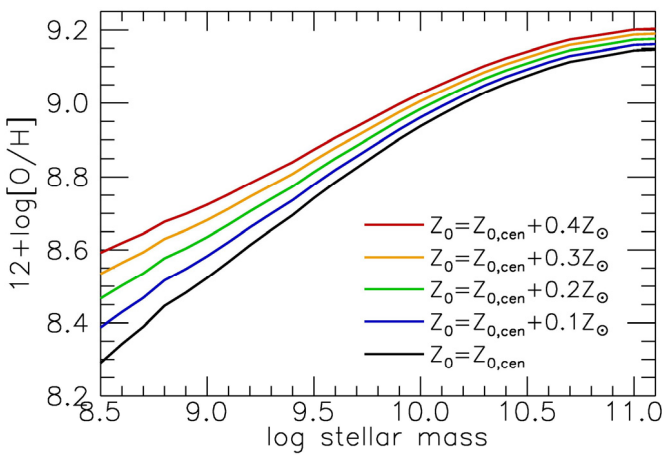

Figure 5. Stellar mass-metallicity relation (MZR) for enriched inflow with different metallicity. Black line shows the observed average metallicity of star-forming central galaxies. Other lines show the MZR for progressively enriched inflow determined by Equation (2).

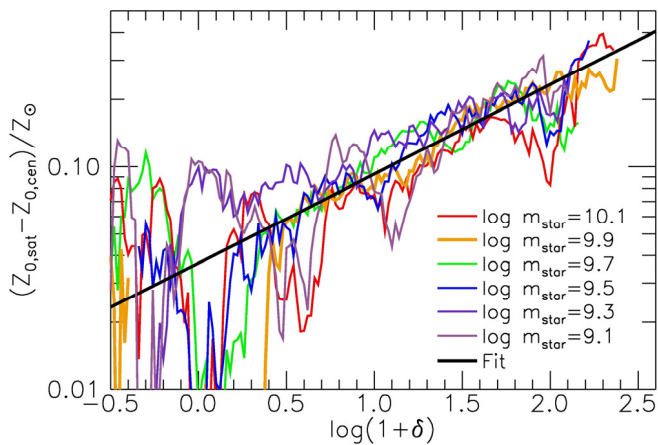

Figure 6. Metallicity of the inflow of star-forming satellite galaxies in units of solar oxygen abundance as function of over-density for different stellar masses. The black line shows the fitted relation to the data for all available stellar masses.

The derived metallicities are shown, in Figure 5, as the color coded lines corresponding to different values of $Z_0$. It is important to note that for a given $Z_0$, the metallicity enhancement is stronger for low mass galaxies and it becomes smaller with increasing mass. This mass-dependent metallicity enhancement is also clearly shown in the top panel of Figure 3. As shown there, at a given over-density, the metallicity enhancement of the star-forming satellites, i.e. the difference between the solid line and dashed line, becomes smaller with increasing stellar mass. Therefore, even if the inflow of the centrals can be metal enriched in dense regions, the central as the most massive and (by extension) the most metal rich galaxy in its group, it will be least metal enhanced in its group.

In the following we attempt to determine the metallicity of the infalling gas as a function of environment, by exploiting our observational results. To determine the metallicity of the infalling gas, we need to know the metallicity enhancement of star-forming satellites compared to the metallicity of star-forming centrals, as function of both stellar mass and environment. This can be directly retrieved from the top panel of Figure 3. As discussed before, at a given stellar mass, satellites and centrals probe different environments and there are few star-forming centrals in the densest regions. Since the metallicity of star-forming centrals is independent of environment, we can use the average metallicity of star-forming centrals (horizontal black dashed lines) as a (good) approximation to the observed gas metallicity of centrals in all environments. Then the metallicity enhancement at a given stellar mass and a given over-density is simply determined by the difference between the horizontal black dashed line and the solid color-coded line.

By inserting the metallicity enhancement at a given stellar mass and a given over-density into Equation (2), the same equation gives the metallicity of the inflow of star-forming satellite galaxies, $Z_{0,sat}$, as a function of over-density and stellar mass. The results are shown in Figure 6. At a given stellar mass and over-density, the uncertainty of the derived $Z_{0,sat}$ is comparable to the uncertainty of the average metallicity of the star-forming satellites shown in the top panel of Figure 3. For clarity, we do not show these uncertainties here in Figure 6. Since at a given over-density the metallicity enhancement of star-forming satellites becomes smaller with increasing stellar mass, at stellar mass above $10^{10.2}$ $M_\odot$ the difference of the average metallicity between star-forming satellites and star-forming centrals becomes undetectable in our data. Therefore these high stellar mass bins are not shown in Figure 6. Future observations with a large sample size and more accurate measurements of metallicity and stellar mass will help to detect such small metallicity difference for the high mass galaxies.

Figure 6 clearly shows that, at a given stellar mass, the metallicity of the inflow of star-forming satellites strongly increases with increasing over-density. Most interestingly, this $Z_{0,sat}$ − overdensity relation is largely independent of stellar mass, at least for the stellar mass range for which we have reliable measurements. The $Z_{0,sat}$ − overdensity relation can be well fitted by a simple power law, given by

$$\log\left(\frac{Z_{0,sat} - Z_{0,cen}}{Z_\odot}\right) = -1.43 + 0.4\log(1+\delta) \quad (3)$$

If the inflow of star-forming centrals is about pristine in all environments as we have assumed before, i.e. $Z_{0,cen} = 0$, then the metallicity of the inflow of star-forming satellites rises from $\sim 0.01 Z_\odot$ in void-like environments (where log(1+$\delta$) <

-1, i.e. ten times below the mean density) to ~ 0.3$Z_\odot$ in cluster-like environments (where log(1+δ) ~ 2, i.e. one hundred times above the mean density) and even up to ~ 0.5 $Z_\odot$ in the core of clusters (where log(1+δ) ~ 3, i.e. one thousand times above the mean density).

Since the gas ejected from galaxies is expected to mix rapidly with the ICM both chemically and thermally (e.g. De Young, 1978), via instabilities, conduction or diffusion, we assume that the metallicity of the inflow of star-forming satellites can represent the metallicity of the IGM. Then Figure 6 can be regarded as the measurement of the metallicity of the IGM in different environments. Encouragingly, these derived metallicities shown in Figure 6 are in broad agreement with observations. The typical cluster gas metallicity is observed to be Z ~ 0.3 $Z_\odot$ (e.g. Arnaud et al. 1994, Mushotzky et al. 1996, Tamura et al. 1996, Mushotzky & Lowenstein 1997). Stocke et al (2007) found metallicity limits of Z < 0.02$Z_\odot$ in the void-like environments at $z < 0.1$. The mean gas metallicity in galaxy filaments at $z \leq 0.15$ has been estimated to be Z ~ 0.1 $Z_\odot$ (Danforth & Shull 2005, Stocke et al. 2006). These observational results therefore indicate that there is indeed a substantial fraction of metals distributed outside of galaxies in the IGM.

Finally, if we assume the $Z_{0,sat}$ – overdensity relation given by the fitting formula in Equation (3) is strictly independent of stellar mass, we can then use Equation (3) and the observed average metallicity of star-forming centrals (black line in Figure 5) to "predict" the metallicities of star-forming satellites at any given stellar mass and over-density via Equation (2). The reproduced metallicities of star-forming satellites are plotted as the black solid lines in the top panel of Figure 3, which are in good agreement with the observed values.

## 5 SUMMARY

In this paper we have explored the dependence of the mass – gas phase metallicity relation for star-forming galaxies on environment, in terms of both over-density and central/ satellite dichotomy in SDSS. By employing the gas regulator model, we have derived the metallicity of the enriched inflow of star-forming satellite galaxies in different environment. The main results of the paper may be summarized as follows.

(1) At a given stellar mass, there is a strong dependence of metallicity on over-density for star-forming satellites, while for star-forming centrals no significant correlation is found. In particular, we find a tight sequence of the average over-density – metallicity relation for the star-forming satellites, spanning the entire observed metallicity range, *independent of stellar mass*. This surprising result implies that there a universal evolutionary path in the average over-density – metallicity plane might apply for all star-forming satellites.

(2) The strong correlation between over-density and metallicity for star-forming satellites suggests that the gas inflow is getting progressively metal-enriched in dense regions. The alternative scenario, of the enhanced metallicity in satellites as a function of environment being ascribed to a higher local production of metals as a consequence of a possibly higher SFR, is very unlikely; indeed, the SFR distribution function at a given stellar mass is universal and depends neither on over-density nor on central/satellite dichotomy. Most models of galaxy evolution assume that the gas inflows are pristine. However, such models should be revised to account for the evidence of (environment dependent) metal enriched inflows obtained by our results. Within this context, one should bear in mind that satellite galaxies (for which this result is clear) account for at least half of the galaxy population.

(3) The inflow of the centrals may be metal enriched by the metal produced and expelled into IGM by their surrounding satellites. However, the independence of metallicity on environment at a given stellar mass for star-forming centrals suggests that such metallicity enhancement is almost negligible, since the amount of metals in the IGM produced by the satellites is too small compared to the metal in the central (as by definition the central galaxy is the most massive and hence the most metal rich galaxy in its group).

(4) We infer that the metallicity of the enriched inflow of star-forming satellite galaxies strongly increases with increasing over-density. The derived $Z_{0,sat}$ – overdensity relation is largely independent of stellar mass and can be well described by a simple power law. If we assume that the metallicity of the inflow of star-forming satellites represents the metallicity of the surrounding IGM, then the implied metallicity of the IGM rises from ~0.01 $Z_\odot$ in void-like environments to ~0.3$Z_\odot$ in cluster-like environments, in broad agreement with observations. This also implies that there is a substantial fraction of metals distributed outside of the galaxies in the IGM.

(5) The observed difference in metallicity between star-forming centrals and star-forming satellites becoming smaller towards high stellar masses can be simply explained by the mass-independent enriched inflow, without the need to involve any mass-dependent environmental effect on metallicity, such as ram-pressure stripping of gas (e.g., Gunn & Gott 1972, Abadi et al. 1999, Quilis et al. 2000), galaxy harassment and tidal stripping (Farouki & Shapiro 1981, Moore et al. 1996).

(6) Environmental effects such as strangulation, ram-pressure stripping and harassment can act to change the gas content of the star-forming satellites and consequently to change their star formation status and can, in principle, contribute to produce the dependence of metallicities on environment. However, we argue that these processes are unlikely to be the main drivers of the observed overdensity – metallicities relation, since these processes would change the SFR of the star-forming satellites in denser regions and therefore contradict the observed universal SFR distribution function at a given stellar mass (as shown in Figure 4).

These new findings greatly extend our understanding of the important role of environment in regulating the metallicity evolution in the galaxy population. The approach presented in this paper can be straightforwardly applied to the high redshift data to study the dependence of galactic metallicity and IGM metallicity on environment at different epochs, and their evolution with cosmic time. This will then help to constrain and better understand the enrichment history of cosmic metals.

These new findings can also be extended to study the dependence of the FMR on environment. Given the fact that at a fixed stellar mass the SFR distribution function is universal and depends neither on over-density nor on central/satellite dichotomy, and that the metallicity of the star-forming centrals is independent of environment, we would expect the FMR of star-forming centrals to be largely independent of environment as well. Likewise, we would expect the FMR of star-forming satellites to be dependent on environment and such dependence would be mainly driven by the dependence of the metallicity on environment, due to the enriched inflow of the star-forming satellites in dense regions. These hypotheses can be directly tested with observations and will be explored in our future work.

**ACKNOWLEDGEMENTS**

We gratefully acknowledge the anonymous referee for comments and criticisms that have improved the paper. We also thank Andrea Ferrara and Romeel Davé for helpful discussions. We are very grateful to Xiaohu Yang and his collaborators for generously providing the DR7 version of their group catalog. We acknowledge NASA's IDL Astronomy Users Library, the IDL code base maintained by D. Schlegel, and the *kcorrect* software package of M. Blanton.